\documentclass[superscriptaddress,showpacs,twocolumn,aps,pre,10pt]{revtex4-1}

\usepackage{amssymb,amsmath,amsfonts}
\usepackage{graphicx,color}
\usepackage{textcomp}
\usepackage[squaren]{SIunits}
\usepackage[english]{babel}
\usepackage{tabularx}
\usepackage[normalem]{ulem}
\usepackage[dvipsnames]{xcolor}

\begin{document}

\title{Colloidal Brazil nut effect in microswimmer mixtures induced by motility contrast}

\author{Soudeh Jahanshahi} 
\affiliation{Institut f\"{u}r Theoretische Physik II: Weiche Materie, Heinrich-Heine-Universit\"{a}t D\"{u}sseldorf, D-40225 D\"{u}sseldorf, Germany}

\author{Celia Lozano}
\affiliation{Fachbereich Physik, Universit\"{a}t Konstanz, Konstanz  D-78457, Germany}

\author{Borge ten Hagen}
\affiliation{Physics of Fluids Group and Max Planck Center Twente, Department of Science and Technology,
MESA+ Institute, and J. M. Burgers Centre for Fluid Dynamics, University of Twente, 7500 AE Enschede,
The Netherlands}

\author{Clemens Bechinger}
\affiliation{Fachbereich Physik, Universit\"{a}t Konstanz, Konstanz  D-78457, Germany}

\author{Hartmut L\"{o}wen}
\affiliation{Institut f\"{u}r Theoretische Physik II: Weiche Materie, Heinrich-Heine-Universit\"{a}t D\"{u}sseldorf, D-40225 D\"{u}sseldorf, Germany}

\date{\today}

\begin{abstract}
We numerically and experimentally study the segregation dynamics in a binary mixture of microswimmers which move on a two-dimensional substrate in a static periodic triangular-like light intensity field. 
The motility of the active particles is proportional to the imposed
light intensity and they possess a motility contrast, i.e., the prefactor depends on the species. In addition, the active particles also experience a torque
aligning their motion towards the direction of the negative intensity gradient.
We find a segregation of active particles near the intensity minima where typically one species is localized
close to the minimum and the other one is centered around in an outer shell. For a very strong aligning torque, there is an exact mapping onto an equilibrium system in an effective external potential that is minimal at the intensity minima. This external 
potential is similar to (height-dependent) gravity, such that one can define effective   ``heaviness'' of the self-propelled 
particles. In analogy to shaken granular matter in gravity, we define a ``colloidal Brazil nut effect'' if the heavier 
particles are floating on top of the lighter ones. Using extensive Brownian dynamics simulations, we identify system parameters
for the active colloidal Brazil nut effect to occur and explain it based on a generalized Archimedes' principle 
within the effective equilibrium model: heavy particles are levitated
in a dense fluid of lighter particles if their effective mass density is lower than that of the surrounding fluid. 
We also perform real-space experiments on light-activated
self-propelled colloidal mixtures which confirm the theoretical predictions.
\end{abstract}

\maketitle

\section{Introduction}
\label{sec:intro}
The physics of active colloidal matter is a rapidly expanding research area on nonequilibrium phenomena.
Typically, active suspensions are composed of self-propelled particles on the micron scale, swimming 
in a fluid at low Reynolds number \cite{Romanczuk2012,Elgeti2015,Cates2012,Bechinger2016,Zoettl_review}. 
The main focus of research has been both on the individual swimming mechanism 
 and on collective effects of many of such microswimmers \cite{Gompper2016}.
The individual swimming speed of a single particle, also called particle motility, is typically of the order of microns 
per second and can be steered externally by various means \cite{Paxton2004,Palacci2010,Volpe2011,Buttinoni2012,Palacci2013JACS,Palacci2013Science,Palacci2014,Moyses2016,Wang2012,Dreyfus2005,Grosjean2015,Steinbach2016,Kaiser2017,Bricard2013,Morin2017}.

Recently, the behavior of microswimmers has been explored in externally imposed motility fields where the swimming speed depends on
 the spatial coordinate \cite{Reichhardt2017}. This not only mimics the chemotactic escape of a living swimming object from toxins or 
its attraction by nutrient gradients \cite{Pohl2014,Saha2014,Liebchen2015,Liebchen2017,Jin2017}, but is also important to steer the directed motion of 
swimmers for specific applications 
such as targeted drug delivery \cite{Gao2012} and nanorobotics \cite{Hong2010}. Various kinds of motility fields have recently been considered
including constant gradients \cite{Hong2007,Ghosh2015}, % more?? 
stepwise profiles \cite{Magiera2015,Grauer2017}, and ratchets \cite{Stenhammar2016,Lozano2016}, as well as time-dependent motility fields \cite{Geiseler2016PRE,Geiseler2017Entropy,Geiseler2017SciRep,Sharma2017}.
In particular, the tunability of the colloid motility by light \cite{Volpe2011,Buttinoni2012,Palacci2013JACS,Palacci2013Science,Palacci2014,Moyses2016,Kummel2013,Buttinoni2013,tenHagen2014NatComm}
provides the opportunity to impose almost arbitrary laser-optical motility fields. When the prescribed light intensity
is proportional to the local motility, a particle will get dynamically trapped in the dark spots where its motility is low \cite{Magiera2015,Grauer2017,Elgeti2}. 

Here we explore a repulsively interacting binary mixture of small self-propelled spherical colloidal particles doped with large ones. The binary mixture of self--propelled colloids is confined to
a two-dimensional substrate in a static periodic triangular-like light intensity field. 
The motility of the particles is proportional to the imposed
light intensity but the prefactor depends on the species. In line with previous experimental 
findings, the light-activated particles also experience a torque
aligning their motion towards the direction of the negative intensity gradient, i.e., swimmers exhibit negative phototaxis \cite{Lozano2016}. This strongly 
favors the dynamical trapping effect near motility minima. 
Using Brownian dynamics computer simulations, we find indeed a demixing of the active particles mixture, where typically one species of particles is
close to the minimum and the other is centered around in an outer shell. 
In the limit of very strong aligning torque, we demonstrate that 
an exact mapping of the nonequilibrium system
onto an equilibrium system is possible. This equilibrium system involves an effective external potential that is minimal at the intensity minima. The external 
potential is piecewise parabolic around the intensity minima. Therefore, it can be understood as an external gravitational potential, where the gravity force depends on the height. Using this analogy, one can define an effective ``heaviness'' of the self-propelled 
particles. Thereby, there is an important link between motility fields of active colloids
and equilibrium sedimentation of passive colloids where a lot of 
theoretical \cite{Lowen2013mixtures,Lowen1998,Torres2007,Wang2008,Lowen2011} and experimental knowledge \cite{Philipse1997,Rasa2004,Piazza1993,Lorenz2009,Brambilla2011}
exists, see Ref.\ \cite{Piazza_review} for a review. 
In analogy to shaken granular matter in gravity \cite{Breu2003,Garzo2008,Godoy2008,lozanoPRL2015,Hong2001,Rosato1987}
and to the sedimentation of colloidal mixtures \cite{Biben1993,Esztermann2004,Zwanikken2005,Spruijt2014,Dijkstra2006,Biesheuvel2005,Kim2015}, we define a ``colloidal Brazil nut effect'' 
if the heavier particles are floating on top of the lighter ones. We identify system parameters
for the colloidal Brazil nut effect to occur and explain it based on a generalized Archimedes' principle \cite{Kruppa2012} within the effective equilibrium model: heavy particles are levitated
in a dense fluid of lighter particles if their effective mass density is lower than that of the surrounding fluid.
 As an aside, another application of the Archimedes' principle has been recently applied to the
lift of passive particles in an active bath \cite{Razin2017}.

We also perform real-space tracking experiments on light-activated colloidal
mixtures. The experimental results agree  quantitatively with the simulation predictions.

The paper is organized as follows: in Sec.\ \ref{sec:model}, we introduce the theoretical model, 
define the colloidal Brazil nut effect, and propose a simple depletion bubble picture to predict the basic physics. Our experiments are described in Sec.\ \ref{sec:exp}. Results from both theory and experiment are presented in the subsequent Sec.\ \ref{sec:results}. Finally, we conclude in Sec.\ \ref{sec:conc}.

\section{Theory}
\label{sec:model}

\subsection{Active Brownian particle model}
\begin{figure}[tbh]
\centering
\includegraphics[width = \columnwidth]{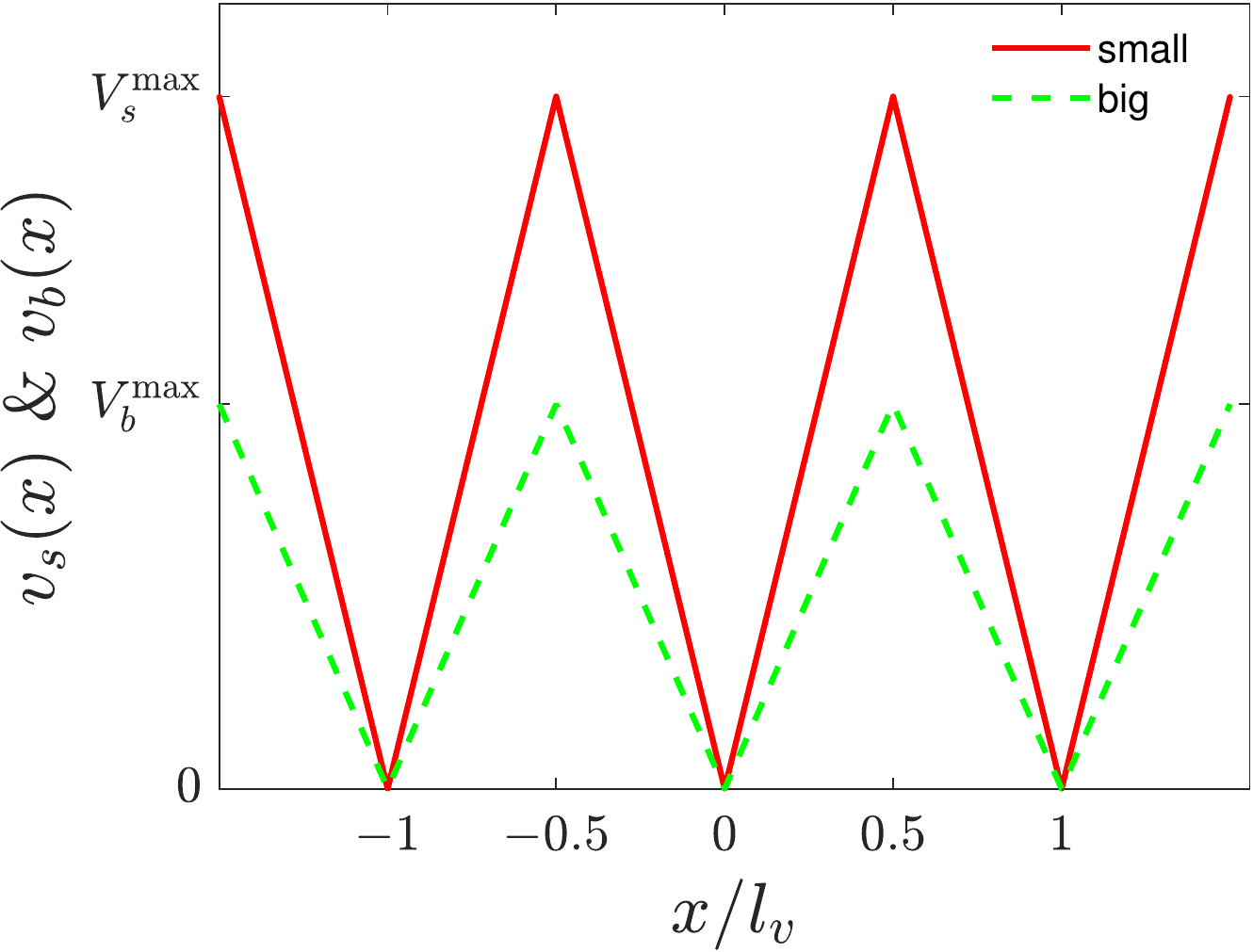}
\caption{\label{vel} Schematic view of the propulsion velocity as a function of $x/l_{v}$ for the two different particle species as originating from a triangular-like light intensity field.}
\end{figure}

We consider an active Brownian particle model for a mixture of big and small 
spheres moving in the two-dimensional $xy$-plane at temperature $T$. The particles have a diameter $\sigma_\alpha$,
where $\alpha = b,s$ (for big and small particles)
is a species index.
The self-propulsion speed of the particles $v_{\alpha}(x)$
depends on their position and is periodic in the $x$-coordinate with a characteristic spacing 
$l_{v}$, but independent of the $y$-coordinate. 
 Having a light motility landscape in mind \cite{Lozano2016}, we assume the same function for both
types of particles except for a different prefactor. In detail, we assume a triangular velocity profile (see Fig.\ \ref{vel}),
for which in one period
\begin{alignat}{1}
v_{\alpha}(x) & =
2\left|x\right|V^{\mathrm{max}}_{\alpha}/l_{v} \quad \mathrm{for}\,\left|x\right|\leq l_{v}/2,
\label{velocityEq}
\end{alignat}
where
  $V^{\mathrm{max}}_{\alpha}$ indicates the maximum propulsion velocity of species $\alpha$.  We consider a large field 
with several  of such velocity grooves, which accommodates $N_\alpha$ particles of species $\alpha$ ($\alpha = b,s$). The system is considered in a rectangular box of edge lengths $L_x$ and $L_y$ with periodic boundary conditions in both directions. Then the partial system densities can 
either be described by areal densities $\rho^{(a)}_\alpha = N_\alpha / (L_xL_y)$ or line densities per wedge $\rho_\alpha = \rho^{(a)}_\alpha l_v$.

The direction of the self-propulsion velocity defines an internal particle orientation degree of freedom which is described by the angle $\varphi$ between the velocity and the $x$-axis.
In addition, there is a  torque aligning the particle orientation along the negative gradient of the motility
field, which leads to an angular velocity $\omega_\alpha$. Note that, in a homogeneous motility field, where $v_{\alpha}(x)$
is constant, this angular velocity obviously vanishes. In general, following our modeling in previous work
\cite{Lozano2016}, the angular velocity $\omega_{\alpha}\left(\varphi,x\right)$ also depends on the $x$ coordinate via 
\begin{equation}
\omega_{\alpha}\left(\varphi,x\right)=\frac{c}{\sigma_{\alpha}}\,v_{\alpha}(x)\,v^{'}_{\alpha}(x)\sin\left(\varphi\right),
\label{torque}
\end{equation}
where $v^{'}_{\alpha}(x)=\frac{dv_{\alpha}(x)}{d\,x}$ denotes the velocity gradient and $c$ is a common prefactor.
Moreover, it was shown \cite{Lozano2016} that the magnitude of the angular velocity scales with the inverse
 of the particle diameter. 

The particles interact via a short-ranged repulsive  Weeks-Chandler-Andersen (WCA) pair potential \cite{WCAPot} 
\begin{alignat}{1}
u_{\alpha\beta}(r) & =
\begin{cases}
u^{\mathrm{LJ}}_{\alpha\beta}(r)-u^{\mathrm{LJ}}_{\alpha\beta}(R_{\alpha}+R_{\beta}) & r\leq \frac{\sigma_{\alpha}+\sigma_{\beta}}{2}\\
0 & r>\frac{\sigma_{\alpha}+\sigma_{\beta}}{2}
\end{cases}
\label{interaction}
\end{alignat}
where $r$ is the interparticle distance, $u^{\mathrm{LJ}}_{\alpha\beta}(r)=4\varepsilon[(\sigma_{\alpha\beta}/r)^{12}-(\sigma_{\alpha\beta}/r)^{6}]$ is the Lennard-Jones potential and the additive repulsion diameters are $\sigma_{\alpha\beta}=2^{-7/6}(\sigma_{\alpha}+\sigma_{\beta})$
($\alpha,\beta \in \{s,b\}$). The repulsion strength  $\varepsilon$ is fixed to $100k_{B}T$, where $k_B T$ is the (effective) thermal energy. 

We describe the center-of-mass positions  of the particles with
\begin{equation}
 \mathbf{r}_{\alpha,k}(t)=\Big(x_{\alpha,k}(t),y_{\alpha,k}(t)\Big)
\end{equation} 
 and their orientations by the unit vectors
 \begin{equation}
\boldsymbol{\hat{u}}_{\alpha,k}=\big(\cos(\varphi_{\alpha,k}),\sin(\varphi_{\alpha,k})\big),
 \end{equation}
 where $\varphi_{\alpha,k}$ are
 the orientational angles. Here, $k \in  \{1,N_\alpha\}$ labels the particles of the same species.

In the active Brownian model, the equations of motion for the translational and orientational degrees of freedom are
coupled overdamped Langevin equations with stochastic noise. In detail, the translational motion 
of the $k^{th}$ particle of species $\alpha$ is governed by
\begin{equation}
\frac{d}{dt}\mathbf{r}_{\alpha,k}={v}_{\alpha}\left(x_{\alpha,k}\right)\boldsymbol{\hat{u}}_{\alpha,k}
+\frac {1}{\gamma_{\alpha}} \mathbf{F}^{\mathrm{int}}_{\alpha,k} +\sqrt{2\frac{k_{B}T}{\gamma_{\alpha}}}\,\boldsymbol{\xi}_{\alpha,k}\left(t\right).
\label{LangevinMain}
\end{equation}
Here, the pairwise repulsive interaction force $\mathbf{F}^{\mathrm{int}}_{\alpha,k}$ is obtained from 
\begin{alignat}{1}
\mathbf{F}^{\mathrm{int}}_{\alpha,k} & =-\boldsymbol{\nabla}_{\alpha,k}\,{\sum_{\beta=b,s}\sum_{i=1}^{N_{\beta}}}' u_{\alpha,\beta}\left(\left|\mathbf{r}_{\beta,i}-\mathbf{r}_{\alpha,k}\right|\right).
\label{interactionForce}
\end{alignat}  The prime symbol indicates the exclusion of the self-interaction, i.e., if $\beta=\alpha$, then $i$ cannot take the value $k$.

The rotational motion of the $k^{th}$ particle of species $\alpha$ is governed by
\begin{equation}
\frac{d}{dt}\varphi_{\alpha,k}\left(t\right)= \omega_{\alpha}\left(\varphi_{\alpha,k},x_{\alpha,k}\right) +   \sqrt{2\frac{k_{B}T}{\gamma_{\alpha}^{r}}}\,\xi_{\alpha,k}^{\varphi}\left(t\right).
\label{Langevinphi}
\end{equation}

 $\boldsymbol{\xi}_{\alpha,k}\left(t\right)=(\xi^{x}_{\alpha,k}(t),\xi^{y}_{\alpha,k}(t))$ and $\xi_{\alpha,k}^{\varphi}\left(t\right)$ describe zero--mean Markovian white noise, with the variance
 \begin{equation}
 \langle \boldsymbol{\xi}_{\alpha,k}\left(t\right) \otimes  \boldsymbol{\xi}_{\alpha',k'} \left(t'\right)\rangle = 
\delta(t-t') \delta_{\alpha \alpha'} \delta_{k k'}  \boldsymbol{1}
 \label{f_st}
 \end{equation}
 and
  \begin{equation}
  \langle \xi^{\phi}_{\alpha,k}\left(t\right)  \xi^{\phi}_{\alpha',k'} \left(t'\right)\rangle = \delta(t-t')\delta_{\alpha \alpha'} \delta_{k k'},
  \label{phi_st}
  \end{equation}
 where $\langle\, \dotsb \rangle$ indicates a noise average, $\otimes$ denotes the dyadic product, and $\boldsymbol{1}$ is the unit matrix.
For species $\alpha$, the translational and rotational friction coefficients are represented by $\gamma_{\alpha}$ and $\gamma_{\alpha}^{r}$, respectively. We neglect hydrodynamic interactions between particles \cite{Liebchen-Intraction2018}.

For spherical
particles with a hydrodynamic diameter $\sigma_{\alpha}$, the friction coefficients are given by $\gamma_{\alpha}=3\pi\eta \sigma_{\alpha}$ and $\gamma^{r}_{\alpha}=\pi\eta \sigma^{3}_{\alpha}$, where $\eta$ is the viscosity of the medium.
The respective short-time translational and rotational diffusion coefficients $D_{\alpha}$ and $D^{r}_{\alpha}$ are characterized by the corresponding friction coefficients, such that 
\begin{equation} 
D_{\alpha}=k_{B}T/\gamma_{\alpha}
\end{equation}
 and 
 \begin{equation}
 D^{r}_{\alpha}=k_{B}T/\gamma^{r}_{\alpha}.
 \end{equation} Thus, for spherical
particles, $D_{\alpha}$ and $D^{r}_{\alpha}$ fulfill 
\begin{equation}
D_{\alpha}/D^{r}_{\alpha}=\sigma_{\alpha}^{2}/3
\end{equation}
when in equilibrium ($\boldsymbol{v}_{\alpha}=0$).

In our active Brownian model,
particles will localize where the self-propulsion velocity is zero, i.e., around $x=nl_v$ with an integer $n$.
There are two reasons for that: first of all, a vanishing mobility implies a larger resting time.
Consequently, even for $c=0$,
 the probability density of an ideal non-fluctuating particle will
 scale as $1/v_{\alpha}(x)$. Fluctuations will lead to an algebraic decay with distance $x$ (when $l_v\to \infty$)
\cite{review_Cate_Tailleur_2015}. Second, and much more importantly here, for $c>0$, 
there is an aligning torque that rotates the particles back such that they travel back 
to  the intensity minimum. The second effect yields exponential localization of an ideal particle
in the groove as a function of $x$ 
when $l_v\to \infty$.

\subsection{Effective equilibrium model}

\begin{figure}[tbh]
\centering
\includegraphics[width = \columnwidth]{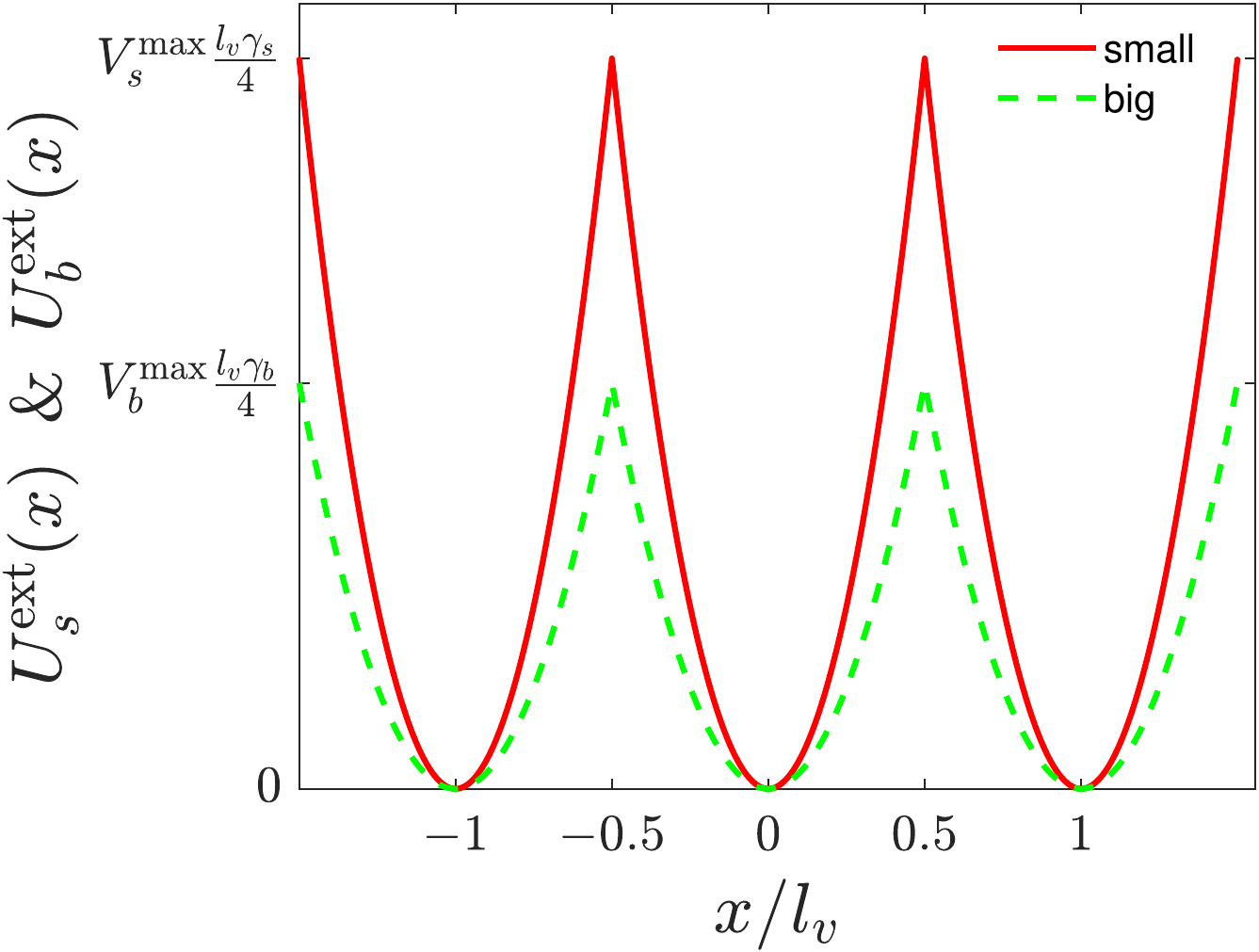}
\caption{\label{extPot} Schematic view of the external potential applied to the particles in the effective equilibrium model as a function of $x/l_{v}$.}
\end{figure}

In the experiments the aligning torque towards the negative gradient of the velocity field 
is strong \cite{Lozano2016} relative to the rotational noise. In this limit, formally achieved by very large prefactors $c$ in  Eq.\ \eqref{torque}, one can neglect the stochastic term in Eq.\ \eqref{Langevinphi}. Then, for all particles,
 the orientation is fixed along the $x$-axis, such that in one period
\begin{equation}
\phi_{\alpha,k}(x_{\alpha,k})=\Bigg\{ \begin{matrix} \pi\quad \quad & \,\,0<x_{\alpha,k}<\frac{l_v}{2}\\ 0\quad \quad & \,\,-\frac{l_v}{2}<x_{\alpha,k}<0
\end{matrix}
\label{OrientationInMap}
\end{equation}
since misalignments are quickly oriented back.
This implies that the self-propulsion velocity in the translational Langevin equation \eqref{LangevinMain}
is directed along the $x$-axis and the resulting term can be derived as a gradient from a ``potential'' function.
This means that the equations of motion in this limit can be written as
\begin{equation}
\frac{d}{dt}\mathbf{r}_{\alpha,k}=\frac {1}{\gamma_{\alpha}}\bigg(\mathbf{F}^{\mathrm{ext}}_{\alpha}\left(x_{\alpha,k}\right)+\mathbf{F}^{\mathrm{int}}_{\alpha,k} \bigg)+\sqrt{2\frac{k_{B}T}{\gamma_{\alpha}}}\,\boldsymbol{\xi}_{\alpha,k}\left(t\right),
\label{map}
\end{equation}
where the external force $\mathbf{F}^{\mathrm{ext}}_{\alpha}(x)$
is a gradient of a potential energy $U_{\alpha}(x)$:
\begin{equation}
 \mathbf{F}^{\mathrm{ext}}_{\alpha}(x)=-\frac{d}{dx} U_{\alpha}(x)\hat{\mathbf{e}}_{x} . 
 \label{effectiveForcex}
 \end{equation}
The equations of motion \eqref{map} describe ordinary Brownian particles --
 with translational coordinates only -- {\it in equilibrium} and 
define our {\it effective equilibrium model}. In general, in analogy to the velocity profile of the active mixture defined via Eq.\ \eqref{velocityEq}, $U_{\alpha}(x)$ is periodic in $x$ with periodicity length $l_v$
and is piecewise parabolic, see
Fig.\ \ref{extPot}. In one period, it is given by
\begin{alignat}{1}
U^{\mathrm{ext}}_{\alpha}(x) &=-\gamma_{\alpha}\int_0^x dx'\,v(x')=-2V^{\mathrm{max}}_{\alpha}\frac{\gamma_{\alpha}}{l_{v}}\int_0^x dx'\, \left|x'\right|\nonumber\\&= V^{\mathrm{max}}_{\alpha}\frac{\gamma_{\alpha}}{l_{v}}x^{2}\quad \mathrm{for}\,\left|x\right|\leq l_{v}/2.
\label{ExtPot}
\end{alignat}

In this equilibrium model, particles would clearly 
accumulate in the minimum of the potential energy, e.g., around $x=0$, in qualitative agreement with
 the 
active Brownian particle model.

\subsection{Definition of the colloidal Brazil nut effect}  
 The Brazil nut effect (BNE) is directly connected to the space-dependent accumulation around the motility minima in the steady state. Information about particle distributions is 
contained in the inhomogeneous one-particle density profiles in the resulting steady state.
For a system homogeneous in $y$-direction, the corresponding density profiles only depend on $x$
and are $l_v$-periodic non-negative functions.
In analogy to equilibrium systems \cite{R_Evans_Advances_in_Physics_1979}, we hence define  line-density profiles
   \begin{equation}
   \rho_{\alpha}(x)=\Big\langle\sum_{i=1}^{N_{\alpha}}\delta(x-x_{\alpha,i})\Big\rangle.
   \label{LineDensity}
   \end{equation}
Here, the brackets $\langle...\rangle$ denote a steady state average and become 
a traditional canonical average in equilibrium (for $v_\alpha(x)\equiv 0$ respectively in the effective equilibrium model).

In order to characterize the Brazil nut effect, we define a spatial extension (or a width) $h_{\alpha}$ 
of the line-density profile in one groove by considering the normalized second moment
     \begin{equation}
     h_{\alpha}=\sqrt{\frac{\int_{-l_v/2}^{l_v/2}\,dx\,x^{2}\,\rho_{\alpha}(x)}{\int_{-l_v/2}^{l_v/2}\,dx\,\rho_{\alpha}(x)}}.
     \label{WidthOfLineDensity}
     \end{equation}
 For sedimentation, this would correspond to an averaged sedimentation height of species $\alpha$.

The effective equilibrium model now helps to define a ``heaviness'' of the particle species. 
The prefactor $\gamma_{\alpha} V^{\mathrm{max}}_{\alpha}\sim\sigma_{\alpha} V^{\mathrm{max}}_{\alpha}$ in Eq.\ \eqref{ExtPot} for the potential energy corresponds to effective heaviness.
Therefore, we define that the big particles are ``heavier'' than the small ones if the following condition is fulfilled:
 \begin{equation}
 \sigma_{b}V^{\mathrm{max}}_{b}> \sigma_{s}V^{\mathrm{max}}_{s},
 \label{effectiveHeaviness}
 \end{equation}
while obviously in the opposite case the smaller particles are heavier than the bigger ones.
By definition a Brazil nut effect occurs if the heavier particles are on top of the lighter ones,
i.e., {\it if the height of the heavier particles is larger than the height of the lighter particles}.
Clearly, there are three
possibilities for that: 
\begin{enumerate}
\item The bigger particles are heavier than the smaller ones, i.e., $\sigma_bV^{max}_b > \sigma_sV^{max}_s$. 
Then a BNE occurs if $h_b>h_s$ We call this situation $\text{BNE}^{(1)}$. 
Conversely, if  $h_b<h_s$, there is a state with the reverse effect, which we refer to as
$\text{reverse BNE}^{(1)}$.
\item  The smaller particles are heavier than the bigger ones, i.e., $\sigma_sV^{max}_s > \sigma_bV^{max}_b$. 
Then a BNE occurs if $h_s>h_b$, this situation is referred to as $\text{BNE}^{(2)}$. 
Conversely, if  $h_s<h_b$, there is a reverse BNE referred to as $\text{reverse BNE}^{(2)}$.
\item The special case when $h_s=h_b$ is termed $\text{no BNE}$.
\end{enumerate}

In conclusion, we have classified the system within a scheme of five possible states: 
$\text{BNE}^{(1)}$, $\text{reverse BNE}^{(1)}$, $\text{BNE}^{(2)}$, $\text{reverse BNE}^{(2)}$, and $\text{no BNE}$. Two of these states 
correspond to a Brazil nut effect where the heavier particles float on the lighter ones. We remark that in the sequel, gravity in our two dimensional system  is directed along the $x$--direction (not along the conventional $y$--direction). So, floating on the top means an outermost layer along the $y$--direction.

\subsection{Depletion bubble picture}
\begin{figure}[tbh]
\centering
\includegraphics[width = \columnwidth]{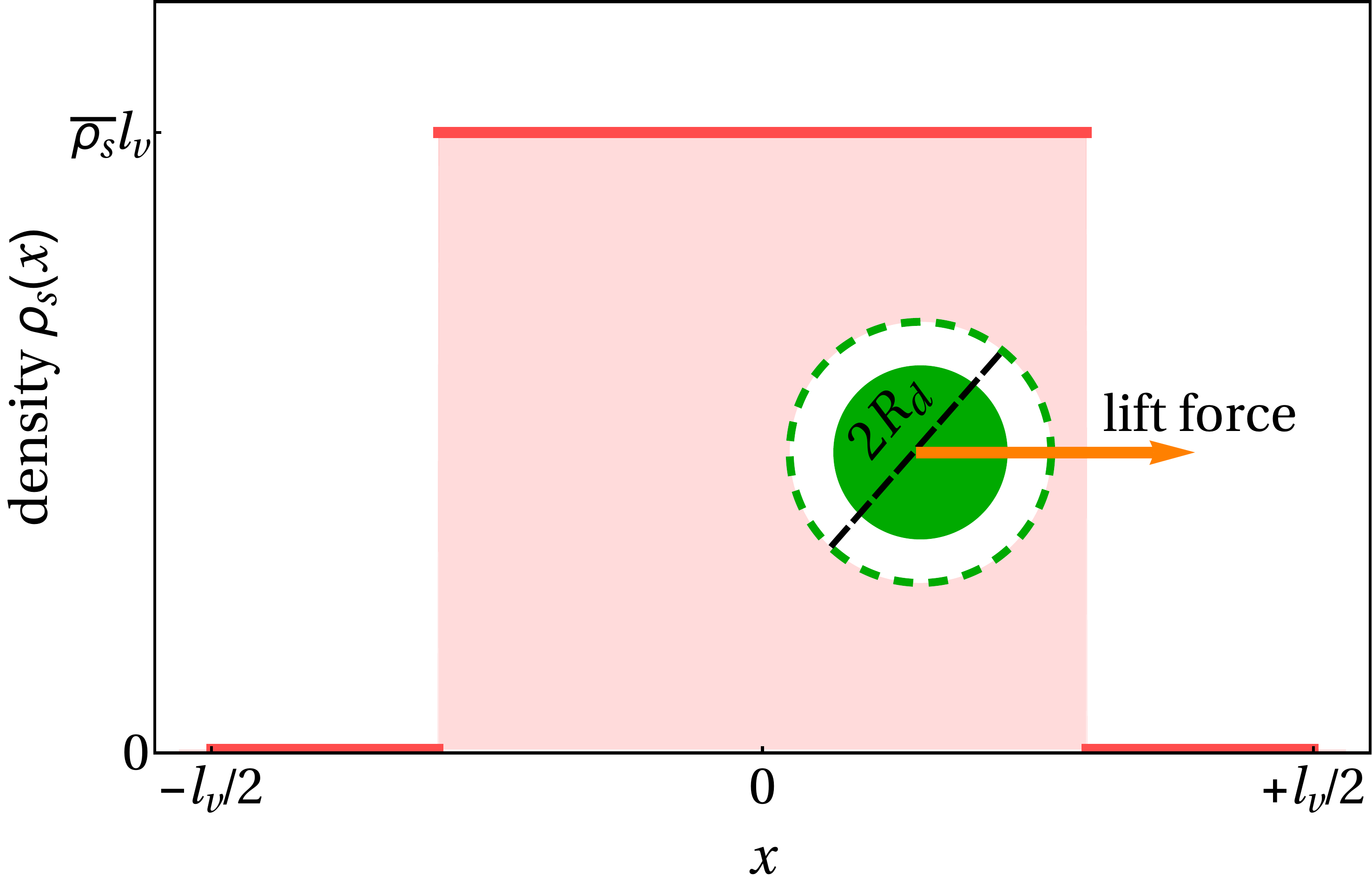}
\caption{\label{DepletionBubbleScheme} Schematic picture of the depletion bubble mechanism. Small particles 
are considered to be uniformly distributed in a fluid block of constant density  $\bar{\rho}_{s}$.
When a big particle delves into this fluid, it will create a
 depletion bubble of radius $R_{d}=(\sigma_{b}+\sigma_{s})/2$. This will result in an equilibrium buoyant 
force according to Archimedes' principle.
}
\end{figure}

We now provide a minimal theory that describes the physics driving the colloidal BNE in terms of a generalized Archimedes' law. 
This approach is based on the effective equilibrium model and was discussed in the context of sedimenting colloidal
 mixtures in Ref.\ \cite{Kruppa2012}. When a big particle excludes
small particles, it creates a bubble or a cavity depleted by small particles. This ``depletion'' bubble is attached
to the big particle and effectively provides a buoyant force which lifts the big particle. For the sake of simplicity, 
let us assume that the density field of the small particles around the groove is piecewise constant, 
i.e., there is a block of fluid at (areal) density $\bar{\rho}_{s}$ (see Fig.\ \ref{DepletionBubbleScheme}).

When a big particle is embedded into this active fluid at a distance $x_b$ from the origin, it will create an encircling
 depletion bubble of radius $R_{d}=(\sigma_{b}+\sigma_{s})/2$ due to the repulsive interactions. 
This bubble is 
attached to the big particles. According to the effective equilibrium model
one can locally apply Archimedes' principle such that the big particle experiences a buoyant 
lift force $F^{\mathrm{buoy}}_{b}$ given by
  \begin{equation}
   F^{\mathrm{buoy}}_{b}(x_b) =\pi R_{d}^{2}\bar{\rho}_{s}F^{\mathrm{ext}}_{s}(x_b),
   \label{buoyantForce}
   \end{equation}
where, from Eq.\ \eqref{effectiveForcex},
\begin{alignat}{1}
F^{\mathrm{ext}}_{s}(x_b) &=-2V^{\mathrm{max}}_{s}\frac{\gamma_{s}}{l_{v}}\left|x_b\right|\quad \mathrm{for}\,\left|x_b\right|\leq l_{v}/2.
\label{F_ext}
\end{alignat}

If the  buoyant lift force dominates the inward effective force (see Eq.\ \eqref{effectiveForcex} again), i.e., if
    \begin{equation}
    F^{\mathrm{ext}}_{b}(x_b) < F^{\mathrm{buoy}}_{b}(x_b)
    \label{buoyancy}
    \end{equation}
   is fulfilled,   the big particles are expelled from the central area of the grooves by the small ones. 
Obviously, the dependence on $x_b$ drops out in Eq.\ \eqref{buoyancy}, such that the condition can be rewritten as
    \begin{equation}
\frac{V^{\mathrm{max}}_{b}}{V^{\mathrm{max}}_{s}} \lesssim \pi\bar{\rho}_{s} (\frac{\sigma_{b} + \sigma_{s}}{2} )^{2}\frac{\sigma_{s}}{\sigma_{b}}.
    \label{criterion}
    \end{equation}

Combined with our previous classification of the Brazil nut effect,
for a given particle heaviness, this approach makes explicit predictions about whether the state
 $\text{BNE}^{(1)}$ occurs or not. However, it requires an input for $\bar{\rho}_{s}$ from simulations and is therefore not fully microscopic. Moreover, this approach only works in the case that the big particles are much more diluted than the small ones.

We finish with two remarks: first of all, correlations will lead to density oscillations in the density profile of the small particles around the big one as discussed in \cite{parola2013JCP}. Second, the converse situation $\text{BNE}^{(2)}$, where a heavy small particle is floating
on a sea of big particles, is also conceivable. This would result from a strongly non-additive large radius $R_d$. A similar depletion bubble picture can be established in this case by interchanging the species indices $b$ and $s$ which we shall, however, not consider further in detail. For more details to the $\text{BNE}^{(2)}$ state, we refer to previous works on passive colloids \cite{parola2013JCP,Piazza2012buoyancy}.

\subsection{Brownian dynamics simulations}
We have solved the equations of motion for the active Brownian model and the effective equilibrium model by using Brownian dynamics computer simulations. In detail, $N_{b}=14$ big and $N_{s}=2068$ small particles were simulated in a periodic square simulation box  with size $L_x = L_y = 102 \sigma_s$, 
which contained 3 complete periods of the motility field, at room temperature. The partial line densities per wedge, $\rho_\alpha$, are thus given by $\rho_s=6.76\,\sigma_s^{-1}$ and $\rho_b=0.046\,\sigma_s^{-1}$.
In terms of a typical Brownian time $\tau=\sigma_s^{2}/D_s$ the time-step $\Delta t$ 
was chosen to be $\Delta t =10^{-5} \tau$. The initial configuration was an ideal gas and the system was equilibrated for an initial time of about $60\tau$. Statistics for the density profiles was gathered during an additional subsequent time window of typically $200\tau$.

In line with the experiments, the maximum velocity of the small particles was fixed at $V^{\mathrm{max}}_{s}=34.5\,\sigma_{s}/\tau$ and the prefactor $c$ was chosen to be  $c=0.6\tau$ \cite{Lozano2016}. 
The simulation results are obtained for different diameter ratios, where $\sigma_s$ has been kept fixed. For each diameter ratio, the maximum velocity of the big particles was varied from $V^{\mathrm{max}}_{b}=0.25V^{\mathrm{max}}_{s}$ to $V^{\mathrm{max}}_{b}=3V^{\mathrm{max}}_{s}$ with steps of $0.25V^{\mathrm{max}}_{s}$. Then, for every value of $V^{\mathrm{max}}_{b}/V^{\mathrm{max}}_{s}$, the occurrence of BNE or reverse BNE has been investigated.

\section{Experiments}
\label{sec:exp}

We experimentally studied concentrated active colloidal mixtures with different size ratios. As small active particles, we used silica spheres of diameter $\sigma_s = \unit{2.7}{\micro\metre}$ half-capped with a carbon layer of thickness $d = \unit{20}{\nano\metre}$. We doped the active suspension with a few large active colloids of diameters $\sigma_b=\unit{13}{\micro\metre}$, \unit{7.75}{\micro\metre}, and \unit{4.96}{\micro\metre}, respectively, while keeping the diameter of small spheres constant. The partial line densities per wedge were approximately $\unit{\rho_s=2.6}{\micro\metre}^{-1}$ and $\unit{\rho_b=0.027}{\micro\metre}^{-1}$  for small and big particles, which are comparable to the line densities used in the simulation ($\unit{\rho_s=2.5}{\micro\metre}^{-1}$ and $\unit{\rho_b=0.017}{\micro\metre}^{-1}$). 

The colloids were suspended in a critical mixture of water and 2,6-lutidine (lutidine mass fraction 0.286), whose lower critical point is at $T_c=\unit{34.1}{\celsius}$. When the solution is kept well below this value, the capped colloids perform an entire diffusive Brownian motion. Upon laser illumination (at wavelength $\lambda$=\unit{532}{\nano\metre}), which is only absorbed by the particle's cap, the solvent locally demixes, and then persistent particle motion is achieved with a constant swimming velocity $v$ which linearly depends on the incident laser intensity \cite{Volpe2011,Buttinoni2012}. For a given cap thickness, independent of the size of the active particles, the same linear dependence $v\propto I$ is observed. Since the propulsion velocity $v$ depends on the absorbed intensity across the particle's cap, the speed can be varied by the cap thickness with the linear dependence $v\propto Id$ \cite{Gomez-Solano2017sci-rep}.

To vary the propulsion velocity in mixtures of big and small particles, our experiments were performed with three different carbon cap thicknesses of the big particles: $d=\unit{5}{\nano\metre}$, \unit{20}{\nano\metre}, and \unit{30}{\nano\metre}. Under our experimental conditions, the maximum velocity of the small species was fixed at $V^{\mathrm{max}}_{s}= \unit{1}{\micro\metre\per\second}$. For the big species, $V^{\mathrm{max}}_{b}$ was varied as follows: $V^{\mathrm{max}}_{b}= \unit{0.25}{\micro\metre\per\second}$, \unit{1}{\micro\metre\per\second}, and \unit{1.5}{\micro\metre\per\second}. The experiments for each combination of big and small particles were repeated between 5 and 20 times to yield good statistical averages.

Periodic triangle-like light patterns were created by a laser line focus being scanned across the sample plane with a frequency of \unit{200}{\hertz}. Synchronization of the scanning motion with the input voltage of an electro-optical modulator leads to a quasi-static illumination landscape \cite{Lozano2016}. Particle positions and orientations were obtained by digital video microscopy with a frame rate of \unit{13}{fps}. The particle orientation was determined directly from the optical contrast due to the carbon cap \cite{LozanoRun-Tumple2018}.  To be more precise, because of the optical contrast between the dark carbon cap and the transparent silica, the angular coordinate $\varphi$ of the active particle can be obtained from the vector connecting the particle center and the intensity centroid of the particle image. The error of this detection is less than $5\%$ as confirmed by comparison with stuck particles whose orientation can be precisely varied using a rotational stage.

\section{Results}
\label{sec:results}

\begin{figure}[tbh]
\centering
\includegraphics[width = \columnwidth]{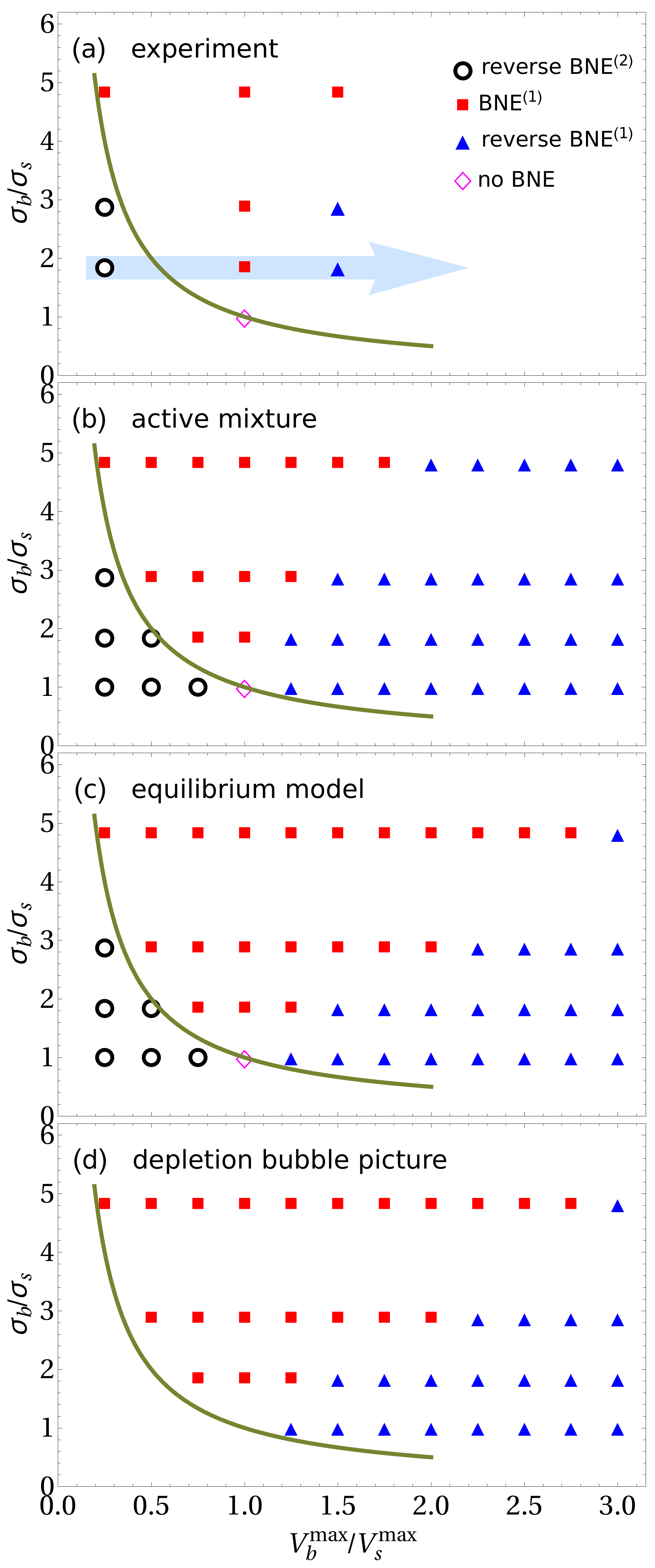}
\caption{\label{phasediag} Occurrence of the Brazil nut effect (BNE) in the parameter space spanned by the motility ratio $V^{\mathrm{max}}_{b}/V^{\mathrm{max}}_{s}$ and the size ratio $\sigma_{b}/\sigma_{s}$ of the binary mixture. Results are shown for: (a) experiment, (b) the active mixture model, (c) the effective equilibrium model, and (d) the depletion bubble picture. The olive green curve indicates the boundary when $F^{\mathrm{eff}}_{b}=F^{\mathrm{eff}}_{s}$. Data are presented for four diameter ratios: $\sigma_{b}/\sigma_{s}=1$, 1.84, 2.87, and 4.82 at fixed $\sigma_s$. More detailed results are shown in the subsequent
Fig.\ \ref{snapshots} for the three parameter combinations highlighted by the light blue arrow in (a).}
\end{figure}

\begin{figure*}[tbh]
\centering
\includegraphics[width = 0.8\linewidth]{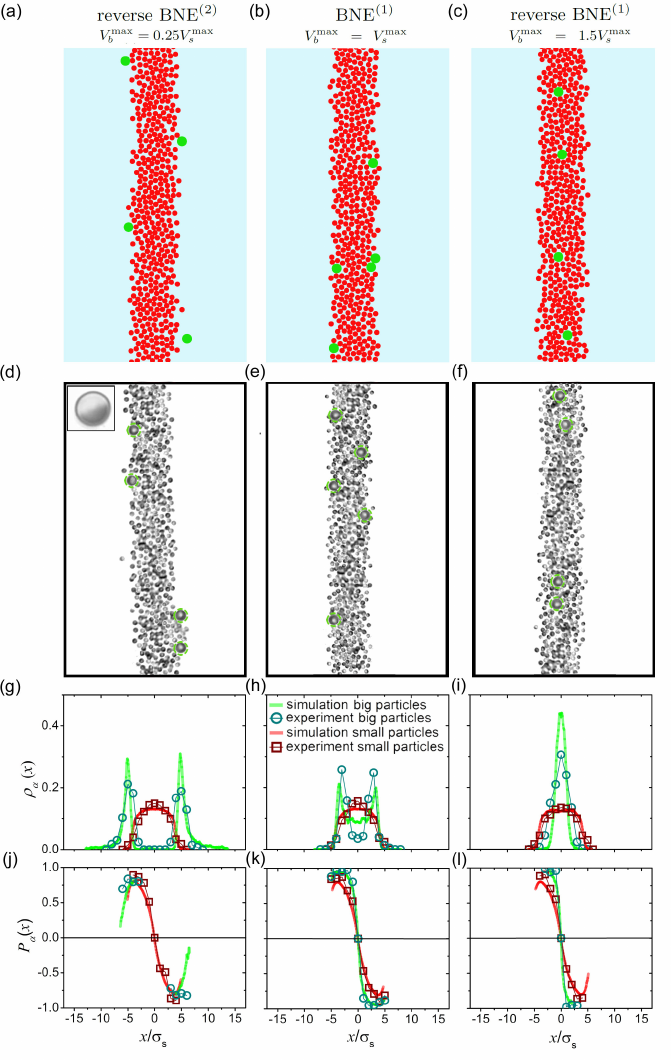}
\caption{\label{snapshots} Comparison of experiment and simulation: (a)--(c) simulation snapshots, (d)--(f) experimental snapshots, (g)--(i) line-density profiles $\rho_{\alpha}(x)$ (defined via Eq.\,\eqref{LineDensity}), and (j)--(l) polarizations $P_{\alpha}(x)$ (introduced in Eq.\,\eqref{polarization}). The results are shown for the $\text{reverse BNE}^{(2)}$ with $V^{\mathrm{max}}_{b}=0.25V^{\mathrm{max}}_{s}$ (first column), the $\text{BNE}^{(1)}$ state with $V^{\mathrm{max}}_{b}=V^{\mathrm{max}}_{s}$ (second column), and the $\text{reverse BNE}^{(1)}$ with $V^{\mathrm{max}}_{b}=1.5V^{\mathrm{max}}_{s}$ (third column). The size ratio is kept constant at $\sigma_{b}/\sigma_{s}=1.84$. Since gravity in our 2D system is along the $x$--direction, floating on the top occurs along the $y$--direction. The inset of (d) shows the microscope picture of a single active particle. }
\end{figure*}

Figure \ref{phasediag} summarizes our main findings in the ($V^{\mathrm{max}}_{b}/V^{\mathrm{max}}_{s}$, $\sigma_b/\sigma_s$)
parameter space of the motility and size ratio of the mixture. The results are shown for (a) the experiment, (b)
the active mixture model, (c) the effective equilibrium model, and (d) the depletion bubble picture. For the considered parameter
span, three different states, namely reverse BNE$^{(2)}$, BNE$^{(1)}$, and reverse BNE$^{(1)}$, were identified
(plus the trivial special case of the no BNE state), see the legend with the different symbols in Fig.\ \ref{phasediag}(a). Remarkably, the topology of the state diagram
is the same in Figs.\ \ref{phasediag}(a)--(d) and there is a quantitative agreement between experiment and theory. As compared to the active mixture model [shown in Fig.\ \ref{phasediag}(b)],
the equilibrium model shows qualitative but not full quantitative agreement. Moreover, the simple depletion bubble pictures is in line
with the equilibrium model.

As expected, the reverse BNE$^{(2)}$ state is favored when $V^{\mathrm{max}}_{b}/V^{\mathrm{max}}_{s}$ is small (i.e., small particles are heavier).
When both species are equally heavy, the crossover from the reverse BNE$^{(2)}$ to the BNE$^{(1)}$  state takes place, as expressed by the
condition $\sigma_{b}V^{\mathrm{max}}_{b}=\sigma_{s}V^{\mathrm{max}}_{s}$, which is shown as the olive green reference
line in Figs. \ref{phasediag}(a)--(d).
In the BNE$^{(1)}$  state, the big particles are heavier but float on the interface. Increasing $V^{\mathrm{max}}_{b}/V^{\mathrm{max}}_{s}$
further leads ultimately to the reverse  BNE$^{(1)}$, as the big particles are getting too heavy to be lifted by the depletion bubble
and sink to the motility minima.
Hence, as the motility asymmetry $V^{\mathrm{max}}_{b}/V^{\mathrm{max}}_{s}$ is increased, the state sequence
$$
\text{reverse BNE}^{(2)} \to \text{BNE}^{(1)} \to \text{reverse  BNE}^{(1)}
$$
is observed. This sequence is reproduced in all of our 4 approaches considered in Figs.\ \ref{phasediag}(a)--(d).
 
 Let us now comment on the comparison between the active mixture and the equilibrium model. The widening of the stability region of the BNE$^{(1)}$ state
 in the equilibrium model can be qualitatively understood in terms of the aligning torque which is strongest in the equilibrium model. If the aligning torque
 is weakened, the demixing is expected to get weaker, favoring the standard reverse  BNE$^{(1)}$ case relative to the BNE$^{(1)}$ state. This
 is indeed observed when comparing Figs. \ref{phasediag}(b) and \ref{phasediag}(c).
 
 The value of $V^{\mathrm{max}}_{b}/V^{\mathrm{max}}_{s}$ where the threshold for the crossover from BNE$^{(1)}$ to reverse BNE$^{(1)}$ happens, grows monotonically with $\sigma_b$/$\sigma_s$.
 This can be explained qualitatively within the depletion bubble picture via the generalized
  Archimedes' law. Assuming that the size $\sigma_s$ and the areal density $\bar{\rho}_s$ of the small species are fixed, the number of small particles excluded
 by a big one grows by increasing the diameter ratio $\sigma_b/\sigma_s$, which results in a stronger buoyant lift force. Based on Eq.\,\eqref{criterion}, the crossover from BNE$^{(1)}$ to reverse BNE$^{(1)}$ is roughly governed by
  \begin{equation}
	\frac{V^{\mathrm{max}}_{b}}{V^{\mathrm{max}}_{s}}\approx\frac{\pi}{4}\bar{\rho}_{s} \sigma_{s}^2\, (\frac{1}{\sigma_{b}/\sigma_{s}} + 1 )^{2}\,\,\frac{\sigma_{b}}{\sigma_{s}}.
     \label{criterion_for_crossover}
     \end{equation}
  The right hand side of the above equation is an increasing function in $\sigma_b/\sigma_s$ (for $\sigma_b/\sigma_s \geq 1$). This implies the crossover from BNE$^{(1)}$ to reverse BNE$^{(1)}$ occurs at larger $V^{\mathrm{max}}_{b}/V^{\mathrm{max}}_{s}$ if the diameter ratio $\sigma_b$/$\sigma_s$ is increased.
 Note that this consideration does not capture the situations on the left hand side of the olive green curve in the parameter space,
 where the big particles are lighter than the small ones, since the depletion bubble picture does not hold here. Last, we remark that we never observe a BNE$^{(2)}$ state for the parameters considered here.
 However, this state is expected to occur in principle in a strongly non-additive binary mixture.
 
 Simulational and experimental snapshots  together with averaged partial density and polarization profiles are summarized
 in Fig.\ \ref{snapshots} for the three states reverse BNE$^{(2)}$, BNE$^{(1)}$, and reverse  BNE$^{(1)}$
 at fixed size asymmetry and increasing  motility asymmetry $V^{\mathrm{max}}_{b}/V^{\mathrm{max}}_{s}$.
 The associated path of parameters is marked by a light blue arrow in Fig. \ref{phasediag}(a). The snapshots clearly indicate
 whether the big particles are floating on the layer of small particles or are confined to the motility minima and therefore
 directly reveal the different states.
 The partial line-density profiles $\rho_{\alpha}(x)$ [defined in Eq.\ \eqref{LineDensity}] reveal a remarkable quantitative
  agreement between experiment and simulation in all three states, see Figs.\ \ref{snapshots}(g)--(i). Most of the deviations are within the
  statistical errors and small systematic deviations  may be attributed to polydispersity and
 hydrodynamic interactions which are neglected in our model.
 
 Finally, we show polarization profiles  in Figs.\ \ref{snapshots}(j)--(l). For a one-component active system under conventional gravity,
 polarization effects have been studied  in theory \cite{Enculescu,Wolff} and experiments \cite{Ginot2}. Likewise, we define the partial
  polarization profiles here as
    \begin{equation}
    P_{\alpha}(x)=\frac{\Big\langle\sum_{i=1}^{N_{\alpha}} \cos(\varphi_{\alpha,i})\,\delta(x-x_{\alpha,i})\Big\rangle}{\rho_{\alpha}(x)}.
    \label{polarization}
    \end{equation}
 Clearly, the polarization is strongly affected by the aligning torque. When a particle crosses the
 motility minimum from left to right (respectively right to left), the torque quickly changes its orientation by \unit{180}{\degree}.
 In the ideal case of instantaneous orientational flips as embodied in the effective equilibrium model, the polarization
 profile would exhibit a sharp kink-like sign function $\text{sgn}(x)$. A finite torque will lead to a smearing of this sign-function, where at the motility minima
 $P_{\alpha}(x=0)=0$ vanishes due to symmetry. If one particle species floats on top of a fluid of the other species,
  there is a non-monotonicity in the polarization, which is well-pronounced for the big particles in Fig.\ \ref{snapshots}(j)
 and for the small particles in Fig.\ \ref{snapshots}(l). This peak in  $P_{\alpha}(x)$ roughly corresponds to the outermost particle layering
 and has its physical origin in the fact that active particles near repulsive walls show a polarization peak in general \cite{Smallenburg2015,Nikola2016}.
  Clearly, the stronger the motility the sharper the polarization profiles.
 Again there is a very good agreement between experiment and simulation,
 supporting earlier findings that the used propulsion mechanism employed in our experiments remains largely unaffected by
 the presence of other nearby particles \cite{Gomez-Solano2017sci-rep,Buttinoni2013}.
 
 \section{Conclusions}
 \label{sec:conc}
 We have presented a systematic study of demixing (or segregation) in binary mixtures of active particles moving on a motility contrast landscape by comparing theory, computer simulations, and experiments. Our findings are based on the strong orientational response of the active particles towards the local minima, which depends on their size and velocity \cite{Lozano2016}. We have shown that the colloidal Brazil nut effect, well established for sedimenting mixtures of passive colloids in the presence of gravity \cite{Kruppa2012}, can also be achieved in mixtures of active colloids being exposed to an inhomogeneous motility field. We define a Brazil nut effect as a situation where the particles of the \emph{heavier} species are floating on the \emph{lighter} ones. Thus, ``heaviness'' is defined by their coupling to the motility contrast. Within this viewpoint, we have considered different parameter combinations for the size and motility asymmetry and, then, mapped out the BNE occurrence.

 We remark that, while active systems consisting of one particle species have been extensively studied in gravitational fields \cite{Palacci2010,tenHagen2014NatComm,Ginot2015,Campbell2017}
 (see also Refs.\ \cite{Dietrich, Tierno} for other aspects of gravity), there are no studies on dense active mixtures under nonuniform motility fields so far. Our theoretical approach can be flexibly applied to other active mixtures regardless of the details of the static external field. This is demonstrated by mapping our active system onto an equilibrium one with a static effective external potential.

 Our qualitative findings can also be exploited for applications. In particular, different kinds of active particles (see Refs.\ \cite{Wysocki,Bain,da_Gama,Wittmann1} for recent studies) can be separated and sorted. This is of particular importance since an inhomogeneous motility field (e.g.\ an external light gradient) can be better controlled than gravity. Moreover, contrary to dynamical separation phenomena (e.g.\ in ratchets \cite{Reichhardt2017}), the separation procedure proposed here is static in the steady state, such that the uppermost layer of floating particles can be removed more easily. Extensions to ternary mixtures are straight-forward and will be considered in future work, where understanding such demixing structures is a prerequisite to create novel materials through active phase separation and self-assembly.

\begin{acknowledgments}
H.L. and C.B. acknowledge funding from the SPP 1726 of the Deutsche Forschungsgemeinschaft (DFG, German Research Foundation); C.B. by the ERC Advanced Grant ASCIR (Grant No.693683).
B.t.H. gratefully acknowledges financial support
through a Postdoctoral Research Fellowship from the Deutsche
Forschungsgemeinschaft -- HA 8020/1-1.
\end{acknowledgments}

\end{document}